\documentclass[pdflatex,sn-apa]{sn-jnl}
\usepackage{color}

\jyear{2022}%

\theoremstyle{thmstyleone}%
\theoremstyle{thmstyletwo}%

\theoremstyle{thmstylethree}%

\begin{document}

\title[Analyzing urban scaling laws in the U.S. over 115 years]{Analyzing urban scaling laws in the United States over 115 years}

\author*[1]{\fnm{Keith} \sur{Burghardt}}\email{keithab@isi.edu}

\author[2,3]{\fnm{Johannes H.} \sur{Uhl}}\email{johannes.uhl@colorado.edu}

\author[1]{\fnm{Kristina} \sur{Lerman}}\email{lerman@isi.edu}
\author[2,4]{\fnm{Stefan} \sur{Leyk}}\email{stefan.leyk@colorado.edu}

\affil*[1]{\orgname{USC Information Sciences Institute}, \orgaddress{\street{4676 Admiralty Way}, \city{Marina del Rey}, \postcode{90292}, \state{CA}, \country{USA}}}
\affil[2]{\orgdiv{Institute of Behavioral Science}, \orgname{University of Colorado Boulder}, \orgaddress{\street{483 UCB}, \city{Boulder}, \postcode{80309}, \state{CO}, \country{USA}}}
\affil[3]{\orgdiv{Cooperative Institute for Research in Environmental Sciences (CIRES)}, \orgname{University of Colorado Boulder}, \orgaddress{\street{216 UCB}, \city{Boulder}, \postcode{80309}, \state{CO}, \country{USA}}}
\affil[4]{\orgdiv{Department of Geography}, \orgname{University of Colorado Boulder}, \orgaddress{\street{260 UCB}, \city{Boulder}, \postcode{80309}, \state{CO}, \country{USA}}}

\abstract{
The scaling relations between city attributes and population are emergent and ubiquitous aspects of urban growth. Quantifying these relations and understanding their theoretical foundation, however, is difficult due to the challenge of defining city boundaries and a lack of historical data to study city dynamics over time and space. To address this issue, we analyze scaling between city infrastructure and population across 857 United States metropolitan areas over an unprecedented 115 years using dasymetrically refined historical population estimates, historical urban road network models, and multi-temporal settlement data to define dynamic city boundaries based on settlement density. We demonstrate the clearest evidence that urban scaling exponents can closely match theoretical models over a century if cities are defined as dense settlement patches. Despite the close quantitative agreement with theory, the empirical scaling relations unexpectedly vary across regions. Our analysis of scaling coefficients, meanwhile, reveals that a city in 2015 uses more developed land and kilometers of road than a city with a similar population in 1900, which has serious implications for urban development and impacts on the local environment. Overall, our results offer a new way to study urban systems based on novel, geohistorical data. 
}



\maketitle

 Cities are expanding at an unprecedented pace, with the majority of the world's population now living in urban environments \cite{UN2019}, and a near-majority in each continent \cite{Montgomery2008}. Cities encompass highly diverse geographies and populations, both in size and composition, yet seemingly robust urban patterns emerge from this complexity~\cite{Zipf1949,Jiang2015_naturalworld,Eeckhout2004,Jiang2011,Rozenfeld2008,Rozenfeld2011,Gabaix2004,Batty2006,Jusup2022,Lobo2020_urban,Reia2022}. These patterns are important not only because of the significance of cities and because they provide a uniform reference frame for comparing cities, but also because they offer a framework to understand emergent phenomena in complex systems. Among the most widely studied examples of urban patterns are scaling relations between features of a city and its population size~\cite{Bettencourt2007,Bettencourt2013,Jusup2022}, of the form
\begin{equation}
    \text{City feature} = \text{exp}(b) \times P^a,
\end{equation}
where $P$ is the city population, $b$ is the \emph{scaling coefficient}, and $a$ is the \emph{scaling exponent}. We define the constant to be $\text{exp}(b)$ such that a log-log plot will have the relation $\text{log}(\text{City feature}) = a\times \text{log}(P) + b$. The universality of these patterns has recently been called into question, however, because 
the scaling exponents appear to vary over time~\cite{Batty2006,Louf2014,Bettencourt2020_longvscross,Keuschnigg2019,Depersin2018}.
In this paper, we analyze the scaling of cities in the conterminous United States (CONUS) over more than a century, from 1900--2015, in order to understand changes in scaling exponents and coefficients over a time period that is longer than in any previous study \cite{Bettencourt2020_longvscross,Keuschnigg2019,Depersin2018}. We define cities based on settlement densities to capture the evolution of city boundaries over time. 
We use dasymetric refinement \cite{Maantay2007,Wei2017,Huang2021} to determine populations within built-up areas. We then utilize novel techniques to capture the developed area, indoor area, building footprint area, and kilometers of roads within these city ``patches.'' 
By capturing scaling over the span of decades, we better understand how urban environments and infrastructure have developed in relationship to their population. Surprisingly, we find that scaling law exponents are stable over time, in contrast to recent research \cite{Depersin2018, Keuschnigg2019}, while scaling law coefficients increase dramatically over time, which implies that a city in 2015 far less dense and contains larger houses than a city with a similar population in 1900.  Overall, our work provides new insights into the persistence of urban patterns, while revealing how different geographies, as defined by e.g., topography, may nonetheless impact scaling relations. This adds nuance to theoretical arguments behind and implications of urban scaling \cite{Bettencourt2013}.

\section*{Results}

\begin{figure*}[tbh!]
    \centering
    \includegraphics[width=1.0\linewidth]{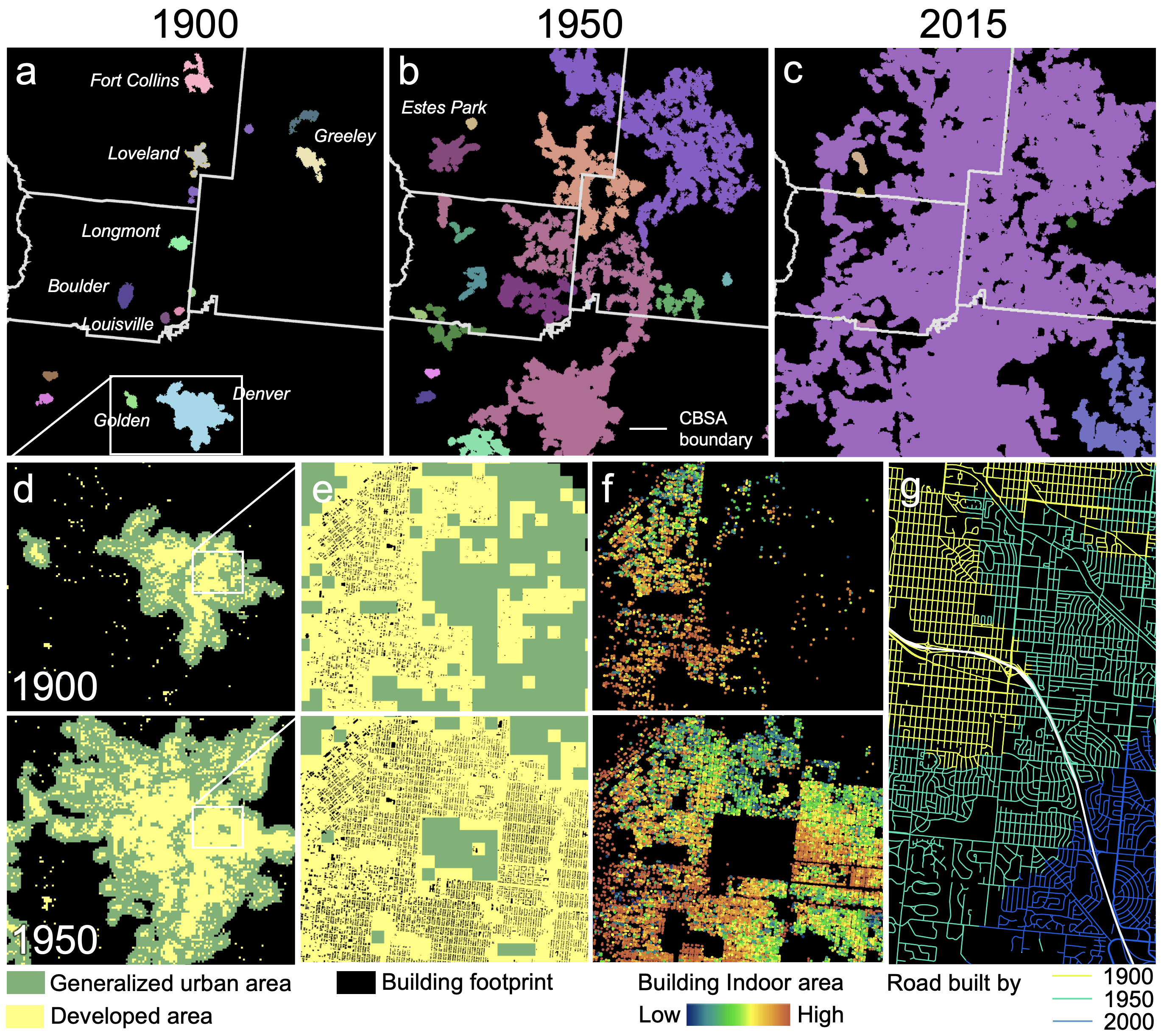}
    \caption{Historical urban delineations used for urban scaling analysis and geospatial data used to measure historical city features for different points in time. Cities in (a) 1900, (b) 1950, and (c) 2015 for the Greater Denver Colorado are, derived from the historical, generalized urban areas dataset \cite{Uhl2022_urbanarea}. Historical (d) developed area (yellow), (e) footprint area (black), (f) building indoor area, and (g) road network at different points in time. Local road networks reconstructed from \cite{Burghardt2021_city}. A modern interstate (whose effect on scaling is negligible) is shown in white.}
    \label{fig:scaling_examples}
\end{figure*}

We recently developed an approach and data infrastructure to measure the long term evolution of settlements \cite{leyk2018hisdac} and road networks for 850 cities within the US since 1900~\cite{Burghardt2021_city}, which adds to the growing literature on reconstructing land use over time \cite{Li2022,Gouj2022}. We apply these data to reconstruct the infrastructure of cities as they grow, as shown in Fig.~\ref{fig:scaling_examples}. 
We treat cities as areas with dense built up settlements that can span large parts of Core-Based Statistical Areas (CBSAs), as shown in Fig.~\ref{fig:scaling_examples}a--c for 1900--2015.  

\begin{figure*}[tbh!]
    \centering
    \includegraphics[width=1.0\linewidth]{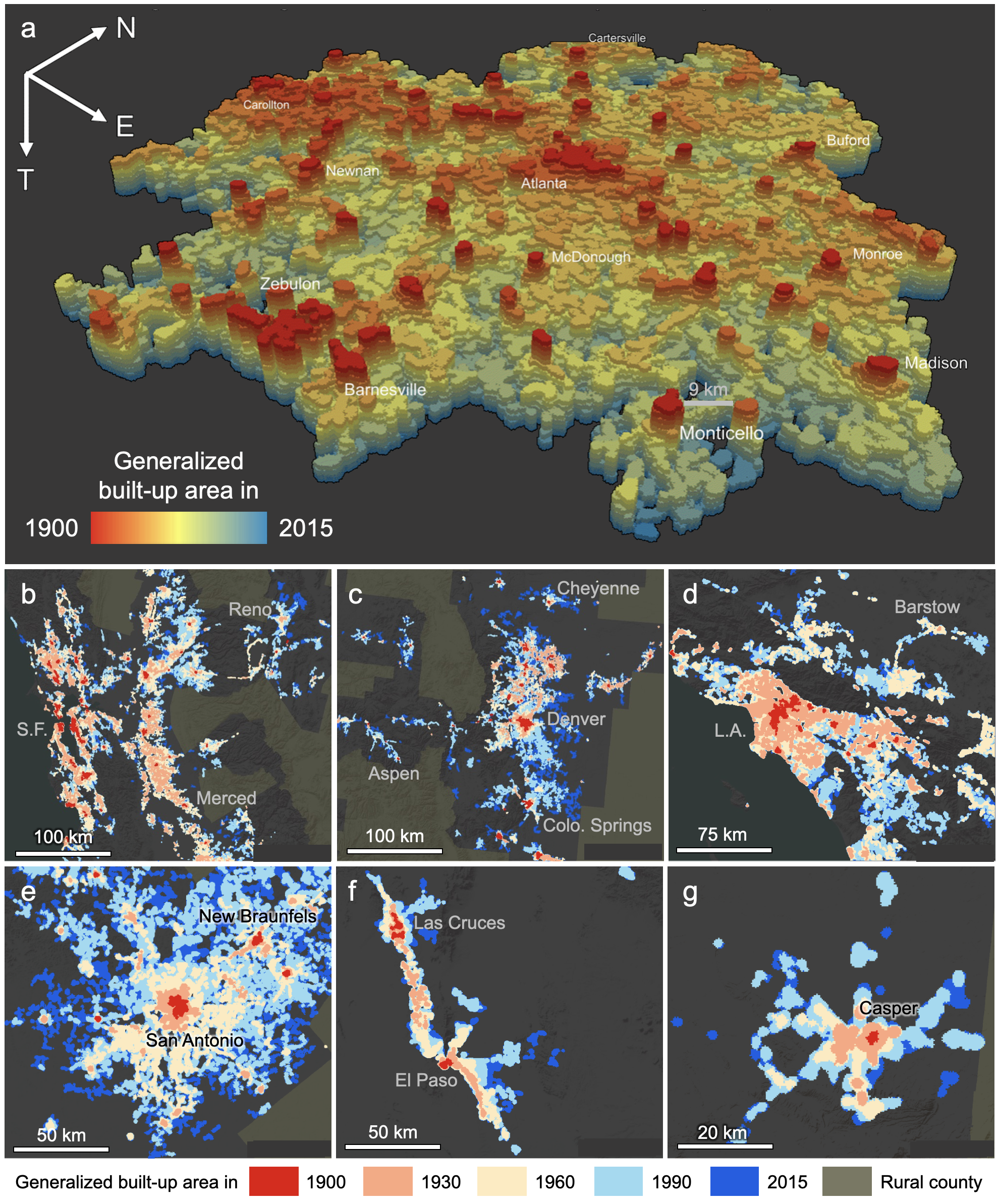}
    \caption{City evolution over time, measured by the Generalized historical built-up area dataset. (a) Generalized built-up area between 1900 and 2015 for Atlanta, Georgia, (b) San Francisco, California, (c) Denver, Colorado, (d) Los Angeles, California, (e) San Antonio, Texas, (f) El Paso, Texas, and (g) Casper, Wyoming.}
    \label{fig:builtup_area}
\end{figure*}

We calculate city features, namely their total developed area, indoor area, building footprint area, road length, and population size (see Materials and Methods for details). Developed area is the land area of developments (yellow area in Fig.~\ref{fig:scaling_examples}d), while building footprint area is the acreage of all buildings within a city boundary (Fig.~\ref{fig:scaling_examples}e). Indoor area (Fig.~\ref{fig:scaling_examples}f) meanwhile represents the total building floorspace across all floors rather than the footprint acreage. Road length is defined as the total kilometers of road found within each city boundary~(Fig.~\ref{fig:scaling_examples}g). We see more examples of built-up areas in Fig.~\ref{fig:builtup_area} for Atlanta, Georgia; San Francisco, California; Denver, Colorado; Los Angeles, California; San Antonio, Texas; El Paso, Texas; and Casper, Wyoming, demonstrating how the developed area spreads from a central location outwards over time.

We then reconstruct historical city populations using gridded HISDAC-US building data \cite{leyk2018hisdac} and historical US Census CBSA populations \cite{census_county_pop_new,census_county_pop_mid,census_county_pop_old}. In many areas of the US, we only know historical populations down to the county-level, therefore to infer populations within arbitrarily defined city boundaries, we find the fraction of settlements within a CBSA that are within the city boundaries and multiply this fraction by the CBSA population that year. The logic behind this methodology is that we observe a strong correlation between the number of buildings and population within a given county (Supplementary Figure S1), therefore the proportional of houses within a city is a good indication of its historical population.  We confirm the accuracy of this method in Supplementary Figure S2 where we compare our population estimation method to alternative methods \cite{Huang2021,Baynes2022}. The city boundaries we define are based on built-up intensity, which is similar to those based on population density \cite{Arcaute2015}, or other metrics \cite{Jiang2010,Jiang2015_evol, Jiang2011}. While there are a number of other ways to define cities \cite{Cottineau2017,parr2007spatial,Small2005}, including those based on workforce and consumption, we use the current method because it is a reasonable proxy of these alternative definitions (people often travel near their home \cite{Gonzalez2009}, making area around a settlement location a good proxy of where people travel and commute). Moreover, our definition is robust enough that we can uncover city boundaries far back in time.


\subsection*{Scaling over time}
\begin{figure*}[tbh!]
    \centering
    \includegraphics[width=1\linewidth]{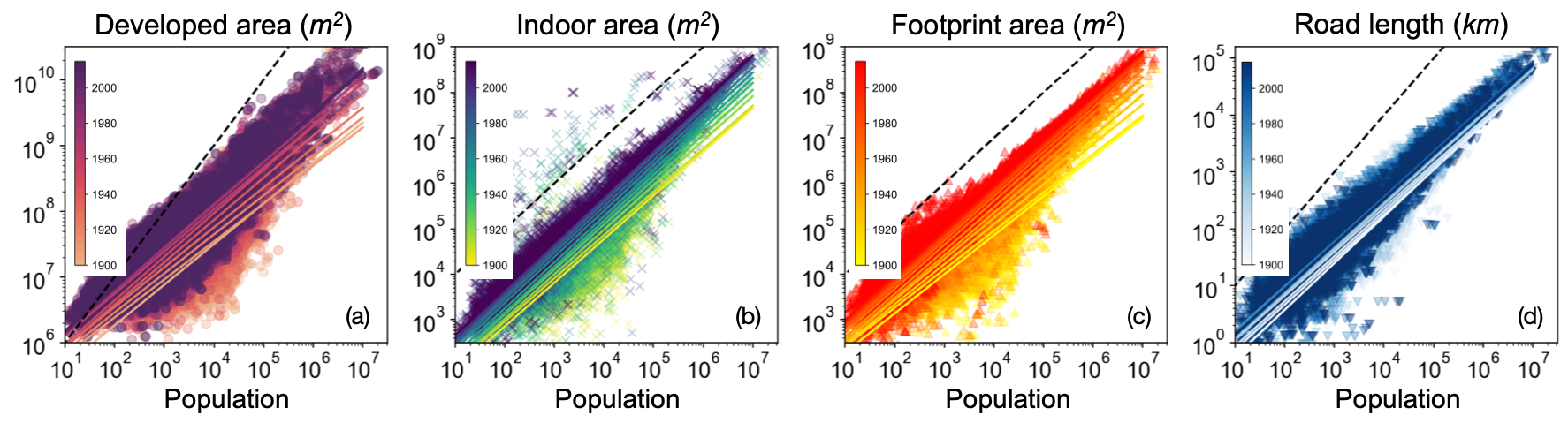}
    \caption{Scaling of infrastructure over time. (a) Developed area, (b) indoor floor area, (c) building footprint area, and (d) road length versus population at decadal intervals from 1900--2010 and 2015.
}
    \label{fig:cross_time}
\end{figure*}

\begin{figure*}
    \centering
    \includegraphics[width=1\linewidth]{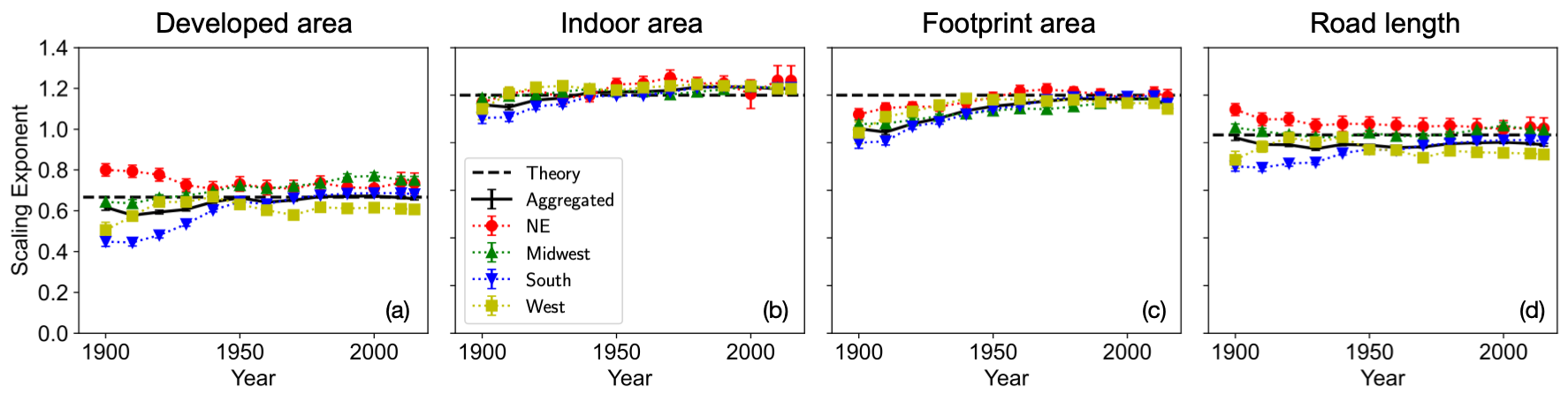}
    \caption{Scaling exponents over time. Exponents for models fitted to (a) developed area, (b) indoor floor area, (c) building footprint area, and (d) road length. Data is separately analyzed by CONUS region: Northeast, Midwest, South, and West. Error bars represent standard errors.}
    \label{fig:ExpOverTime}
\end{figure*}

We show scaling relations in Fig.~\ref{fig:cross_time}, in which the slope (scaling exponent) of each scaling relation is relatively stable, while the y-intercept of the scaling relations (scaling coefficient) tends to increase over time. We quantify these findings as shown for scaling exponents in Fig.~\ref{fig:ExpOverTime} and scaling coefficients in Fig.~\ref{fig:CoefOverTime}. These results are robust to variations in data quality, as shown in Supplementary Figures S3 \& S4. Cities produce a consistently sublinear scaling between developed area or road length and population, and are in close agreement with a theoretical 2/3 law for area, and a 5/6 law for road length \cite{Bettencourt2013}, where the theory is based on cities being defined by where people travel rather than settlement density. The different ways to define cities implies that we should not necessarily expect close agreement with the theory, yet its agreement is tantalizing. One important finding that has not been reported in previous analyses, however, is a consistent linear relation between building footprint area or indoor floor area and population, meaning that large and small cities have similarly sized homes on average.
\begin{figure*}[tbh!]
    \centering
    \includegraphics[width=1\linewidth]{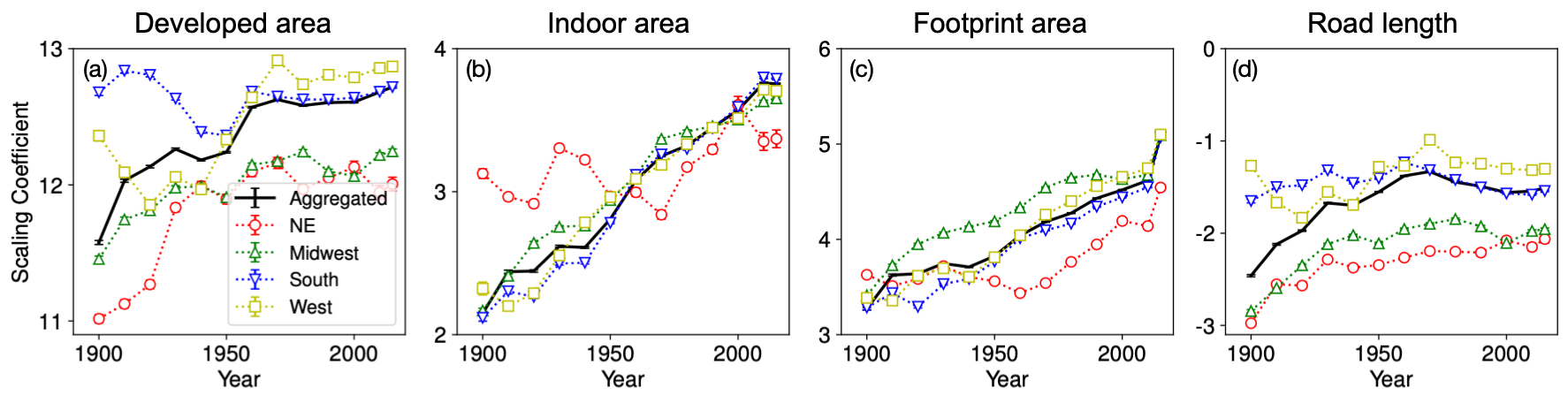}
    \caption{Scaling coefficients over time. Coefficients for models fitted to (a) developed area, (b) indoor floor area, (c) building footprint area, and (d) road length. Data is separately analyzed by CONUS region: Northeast, Midwest, South, and West. Error bars represent standard errors.}
    \label{fig:CoefOverTime}
\end{figure*}

We also notice in Fig.~\ref{fig:CoefOverTime} that the scaling coefficients vary strongly over time, most notably for developed, indoor, and building footprint areas. Given the stable scaling exponent this finding implies the developed and housing area per capita has increased over time, which is consistent to trends observed in the literature \cite{Moura2015}. We independently confirm this finding using US Census data on newly built homes since 1978 (Supplementary Figure S5), in which we find the average area of houses sold has consistently increased until 2010--2015. 

We also notice in Fig.~\ref{fig:CoefOverTime} that developed area continues to increase over time, with one of the largest jumps occurring between 1950--1960. The coefficient change of $\sim 0.5$ implies developed area per capita increased by around 65\%. This finding is consistent with literature on post-WWII housing boom \cite{Moffatt2021}, and white flight \cite{Frey1979}, which have contributed to a greater build-up of housing in suburbs and less-developed areas, after controlling for city population. Finally, we notice that the road length has recently decreased for cities even as their developed area stagnates. This might imply that cities have more recently become more sparse, but more work needs to be done to confirm these findings. Scaling coefficients are therefore not universal and vary strongly over time. The increase of coefficients also implies a greater amount of land used for contemporary development than in a comparable city in the year 1900.

\section*{Spatial Variations in Cross-Sectional Scaling}

In order to test spatial and temporal variation of scaling laws, we ran our analysis for the different regions within the CONUS. We show an example of this in Fig.~\ref{fig:ExpOverTime}, in which we compare scaling laws based on all cities (``aggregated'') to scaling laws based on cities within the Northeast, Midwest, South, and West. Scaling law exponents consistently differ between regions; exponents in 2015 are statistically significantly different across regions (z-score p-values$<0.001$). Surprisingly, we find that the Northeast has consistently higher scaling exponents while the South and West have lower exponents. These lower exponents imply the South or West may have a greater economy of scale because their area or road length per capita is smaller for larger cities, thus they better (or more efficiently) utilize the infrastructure they have.

Furthermore, we also notice in Fig.~\ref{fig:CoefOverTime} that the South and West have the highest coefficients, which implies that cities take up more space in these regions compared to similarly-sized cities in other regions. The most compact cities are therefore in the North-East, which have less developed area (and subsequently fewer roads), as well as smaller building footprints, than other regions. While North-Eastern cities were compact, they also initially had higher indoor area (e.g., larger houses). This did not change drastically over time, in contrast to other regions, leading North-Eastern cities to ultimately have lower indoor area than similar cities in other regions by 2015.

\section*{Discussion and Conclusions}

In this study, we have tested the robustness of scaling laws over space and time. Our empirical definition of cities accounts for cities expanding and evolving, thus capturing when cities expand beyond and across administrative boundaries. In agreement with theoretical models \cite{Bettencourt2013}, we find scaling law exponents do not change significantly over time. Scaling exponents, however, vary across space, possibly due to topographic constraints such as rivers or mountains, which can affect urban density \cite{Angel2017}, but also other constraining factors. This builds on some recent research that finds temporally stable but spatially varying emergent urban patterns \cite{Barthelemy2019}, such as in population dynamics \cite{Reia2022}. 

The long time window we analyze allows us to explore an under-appreciated component of scaling: the scaling coefficient (in contrast to the scaling exponent) shows an increase over time for most features, as seen in Fig.~\ref{fig:ExpOverTime}. Therefore, developed area, indoor floor area, building footprint area, and road length per capita are increasing on average. This reduces the amount of otherwise undeveloped land, which complicates resource use and harms the local environment, including water quality and wildlife habitat \cite{glaeser2004sprawl,lawler2014projected,cumming2014implications}. While previous work has found the developed land base has recently become denser \cite{Bigelow2022}, our work implies that a city 100 years ago was still much more densely developed than the same city would be today. This reduced density is consistent with previous work on the loss of dense urban environments \cite{Angel2017}. One reasonable hypothesis is that the reduced urban density is a function of increased car use throughout the century \cite{Schafer2017}, but we can only draw associations not cause and effect from these data, and previous work found car ownership can also be correlated with increasing urban density \cite{Angel2017}. Therefore, more research is needed to better understand this trend. 

These results have important consequences for the urban sciences. First, urban systems need to be studied based on historical data to fully understand their development trajectories. While previous work has made impressive advances in understanding historical urban development \cite{Ortman2020,Lobo2022_hunter,Lobo2020_scaling,smith2021}, there has been a lack of analysis on scaling laws for the same cities over time. Our analysis exploits recently created historical urban data, which allows scaling comparisons to be performed for the same cities over a century. More analysis of this type, especially for superscaling variables like income \cite{Keuschnigg2019,Bettencourt2007} need to be explored. Finally, regional variations need to be explored. While universal laws are idealized explanations of urban science, regional variations can give insights into region-specific conditions. This can improve the interplay between urban science and policy-makers who focus on region- and city-specific problems. 

There are certain limitations in the data we analyze. For example, built years are missing for many records, and the historical data exhibits survivorship bias \cite{Boeing2020multi}. Furthermore, we assume that roads were constructed at approximately the same time as nearby buildings and that population is proportional to the number of buildings within a given city. We evaluated the reliability of such assumptions to the best of our ability, but more research is needed to test and improve upon them. For example, it is important to uncover new ways to approximate city populations as well as city features over time. In addition, our analysis of city-wide features belies variations within cities that need to be further analyzed and modeled \cite{Xue2022}. Finally, there are several other features of cities, including economic indicators or road topology \cite{scheer2001anatomy,Geiger2020,Xue2022}, that could be explored in future work.

\section*{Materials and Methods}
 \textbf{Historical urban boundary modeling.} We modeled historical urban boundaries based on the historical, generalized built-up areas (GBUA; \cite{Uhl2022_urbanarea}), which are based on the Historical Settlement Data Compilation for the US (HISDAC-US; \cite{leyk2018hisdac}). Specifically, the GBUA dataset has been created by generating focal built-up density layers using gridded data on built-up areas \cite{databua2020}
 available in a grid of cells of 250 m x 250 m, in 5-year intervals from 1900 to 2010, using a circular focal window of 1km radius. From these built-up density layers, only areas of $\ge5$\% built-up density were retained, to exclude low-density, scattered rural settlements. The remaining grid cells (delineating higher density settlements) were then segmented to obtain vector objects for each group of contiguous higher density grid cells (called ``patches''). For each patch, we calculated the total number of buildings contained in it (using historical building counts from HISDAC-US i.e., the Built-Up Property Location (BUPL) layers, \cite{databupl2020}) using a GIS-based zonal statistics tool, and ranked the patches by the number of buildings. Finally, we retained the upper 10\% of these ranked patches, representing the high-density patches only. This process was done separately within each 2010 CBSA polygon \cite{CBSA}. Lastly, the patches in all CBSAs were dissolved based on adjacency only, creating natural, contiguous built-up patches i.e., potentially extending across CBSA boundaries. This process was done for each point in time. 
 
\textbf{Historical population modeling.} We acquired historical population counts at the US county level (boundaries of 2010, National Historical Geographic Information System \cite{NHGIS}) and aggregated the county populations by CBSA to obtain population time series for each 2010 CBSA boundary, for each decade from 1900 to 2010 (from the decennial censuses) and for 2015 (from the American Community Survey) \cite{census_county_pop_old,census_county_pop_mid,census_county_pop_new}. We spatially refined these CBSA-level population counts, for each point in time. To do so, we carried out a dasymetric refinement \cite{Maantay2007,Wei2017,Huang2021} by constraining the population estimates reported for the CBSAs to the generalized built-up areas of the respective year, within each CBSA, and assigning population counts to each patch~proportionally to the number of built-up property records \cite{databupr2020}. The BUPR variable counts the number of built-up cadastral properties within each grid cell, and thus, approximates the spatial distribution of built-up structures, or cadastral parcels, or address points, which have been used for dasymetric modeling in previous work (e.g., \cite{Maantay2007,Wei2017,Huang2021,zandbergen2011dasymetric,tapp2010areal,yin2015disaggregation}). Moreover, previous work has revealed strongly linear correlations between the number of houses within a spatial unit and its population over long time periods (\cite{Leyk2020}; see also Supplementary Figure~S1). Our historical, dasymetrically refined population estimates per city represent unique, long-term depictions of urban populations at fine spatial grain, and are in strong agreement with more sophisticated population estimations, shown in Supplementary Figure S2.  Finally, we calculated the refined population estimates for the cities. 

\textbf{Historical built environment characteristics.} We used three variables to historically characterize the built environment within each urban area representation: (a) developed area (measured by the number of built-up grid cells from the HISDAC-US Built-Up Area (BUA) layers, per CBSA polygon, for each year; (b) indoor area (measured by the built-up intensity layer from HISDAC-US; Built-Up Intensity (indoor area) \cite{data_bui2018}, which reports the total indoor area of all builings within a grid cell, per year); and (c) building footprint area per grid cell and year \cite{BUFA}. To obtain the latter, we spatially integrated cadastral data (containing construction year information) available at discrete, point-based locations from the Zillow Transaction and Assessment Dataset (ZTRAX; \cite{zillow2016}), with vector data representing each building footprint in the US from Microsoft \cite{MBF2020}. This integration assigns the thematic information from each ZTRAX record to the closest building footprint and thus, allowed us to merge the attribute richness of ZTRAX with the fine spatial detail of the USBuildingFootprints dataset. Using these integrated spatial data, we were able to stratify the building footprint data by their construction year and calculated the aggregated building footprint area within each grid cell as defined by the HISDAC-US 250 m x 250 m grid. For each of the gridded surfaces measuring developed area, indoor area, and building footprint area, for each year, we calculated zonal sums in a GIS, within each patch~of the respective year. These zonal sums were then further aggregated to a city based on the recorded spatial relationships.

\textbf{Historical road network modeling.} In order to model historical road network densities, we followed a method proposed in Burghardt et al. \cite{Burghardt2021_city}. Specifically, we clipped contemporary road network vector data from the National Transportation Dataset \cite{NTD2020} to the historical urban delineations from the generalized urban area dataset for each year, and calculated the sum of the road network segment lengths per patch~and year. We then aggregated these road network length estimates to the city based on the recorded spatial relationships (cf. \cite{Burghardt2022_roadstats}). We show scatter plots of relationships between urban characteristics in Supplementary Figure S6 for 1900 and 2015. 

\textbf{Multi-temporal, cross-sectional scaling analysis.} To find scaling laws between population estimates and the metrics describing the historical urban built environment and road network, we took the log of the feature versus the estimated population within the city~and computed the least-squares fit of these trends, for each point of time. The slope of this line corresponds to the exponent of the scaling law. Namely, let $Y$ be the city feature (e.g., developed area), and $P$ be the population. We fitted $Y = A\times P^{B}$ as the linear fit of $\text{log}(Y) = \text{log}(A) + B\times \text{log}(P)$. The y-intercept is $\text{log}(A)$, while the exponent, $B$ describes the slope. For regional stratification of our scaling results, we used geospatial data on Census Regions in the US \cite{CensusRegions}. 

\textbf{Uncertainty quantification, cross-comparison and validation.}
Data Coverage and quality. Approximately 10\% of US counties have low geographic coverage, or low completeness of construction date information (i.e., temporal coverage) in HISDAC-US and the underlying ZTRAX data. These completeness issues have been extensively studied in \cite{Boeing2020multi,Burghardt2021_city,leyk2018hisdac}). We excluded CBSAs affected by these incompleteness issues from our analyses using a geospatial completeness threshold of 40\% and temporal coverage of 60\%. Out of 857 CBSAs, 647 were complete enough for analysis. Settlement density-based city patches, however, can span CBSAs with both high and low temporal coverage or geospatial completeness, therefore we created a weighted average estimate of the temporal coverage or geospatial completeness for each city. The weight was given by the number of patches~within each CBSA that make up a given city. We then thresholded these weighted values to have temporal coverage above 60\% and geospatial completeness above 40\%. 

We conducted a sensitivity analysis showing that our results are largely robust to the choice of this exclusion threshold (Supplementary Figures S3 \& S4)
. Given the robustness of our results, we do not believe data completeness strongly affects our conclusions. Moreover, the BUPR, BUPL, BUA, and indoor area layers have been validated extensively in prior work \cite{Uhl2021_datasets} and show high levels of coherence to other historical measurements of population and building characteristics. Effects of survivorship bias (i.e., the limitation of the HISDAC-US to measure urban shrinkage and historical changes in the built environment) are assumed to be of minor nature (e.g., \cite{Burghardt2021_city,boeing2020off,barrington2015century,Meijer2018}). Historical road network models have been cross-compared to other models and datasets in \cite{Burghardt2021_city} reporting that there is sufficient consistency between these products.

Code for all our analysis can be found at:\url{https://github.com/KeithBurghardt/urbanscaling}.

\section*{Data Availability}
The data is available in the following references: US regions \cite{CensusRegions}, population \cite{census_county_pop_old,census_county_pop_mid,census_county_pop_new}, BUPR \cite{databupr2020}, BUPL \cite{databupl2020}, BUA \cite{databua2020}, footprint area \cite{BUFA}, and indoor area \cite{data_bui2018}, road networks \cite{Burghardt2022_roadstats}, and patch boundaries \cite{Uhl2022_urbanarea}. SI data for house sizes since 1978 are from \cite{HouseArea}. In these data, references to patches are density-based settlements within a CBSA. Merging any neighboring cross-CBSA patches (using \cite{Uhl2022_urbanarea}) creates the city features we analyze in this paper.

\section*{Competing Interests}
The Authors declare no competing financial or non-financial interests.

\section*{Author Contributions}
K.B., J.U., K.L., and S.L. designed research; K.B. and J.U. performed analysis; K.B., J.U., K.L., and S.L. wrote the paper.

\section*{Acknowledgements}
 Partial funding for this work was provided through the National Science Foundation (awards 1924670 and 2121976) and the Eunice Kennedy Shriver National Institute of Child Health and Human Development of the National Institutes of Health under award number P2CHD066613. The content is solely the responsibility of the authors and does not necessarily represent the official views of the National Institutes of Health. Moreover, Safe Software, Inc., is acknowledged for providing a Feature Manipulation Engine (FME) Desktop license used for data processing.

\setcounter{figure}{0}

\makeatletter 
\renewcommand{\thefigure}{S\@arabic\c@figure}
\makeatother

\begin{figure*}[tbh!]
    \centering
    \includegraphics[width=0.8\linewidth]{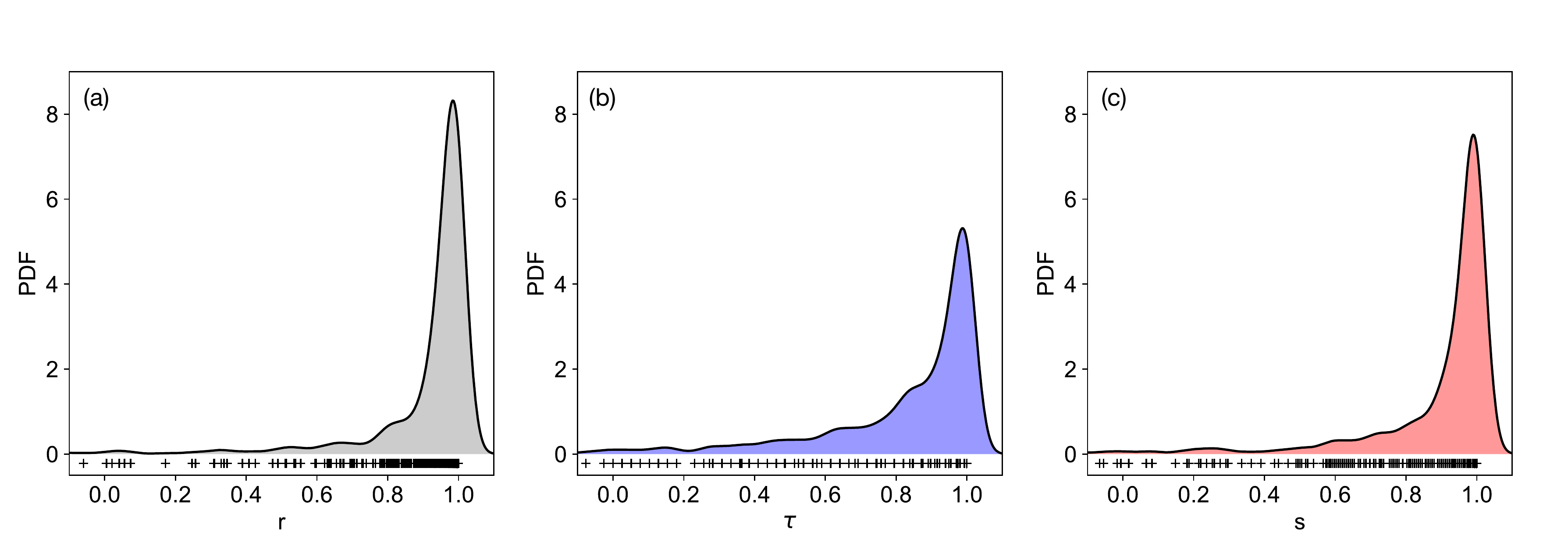}
    \caption{Correlations between number of building footprints and CBSA population over time. Distributions are for each CBSA with temporal completeness over 60\% and geospatial completeness over 40\% from 1900 and 2015 (647 CBSAs). (a) Pearson correlation, (b) Kendall's Tau, and (c) Spearman rank correlation.}
    \label{fig:pop_correl}
\end{figure*}

\begin{figure*}
    \centering
    \includegraphics[width=0.95\linewidth]{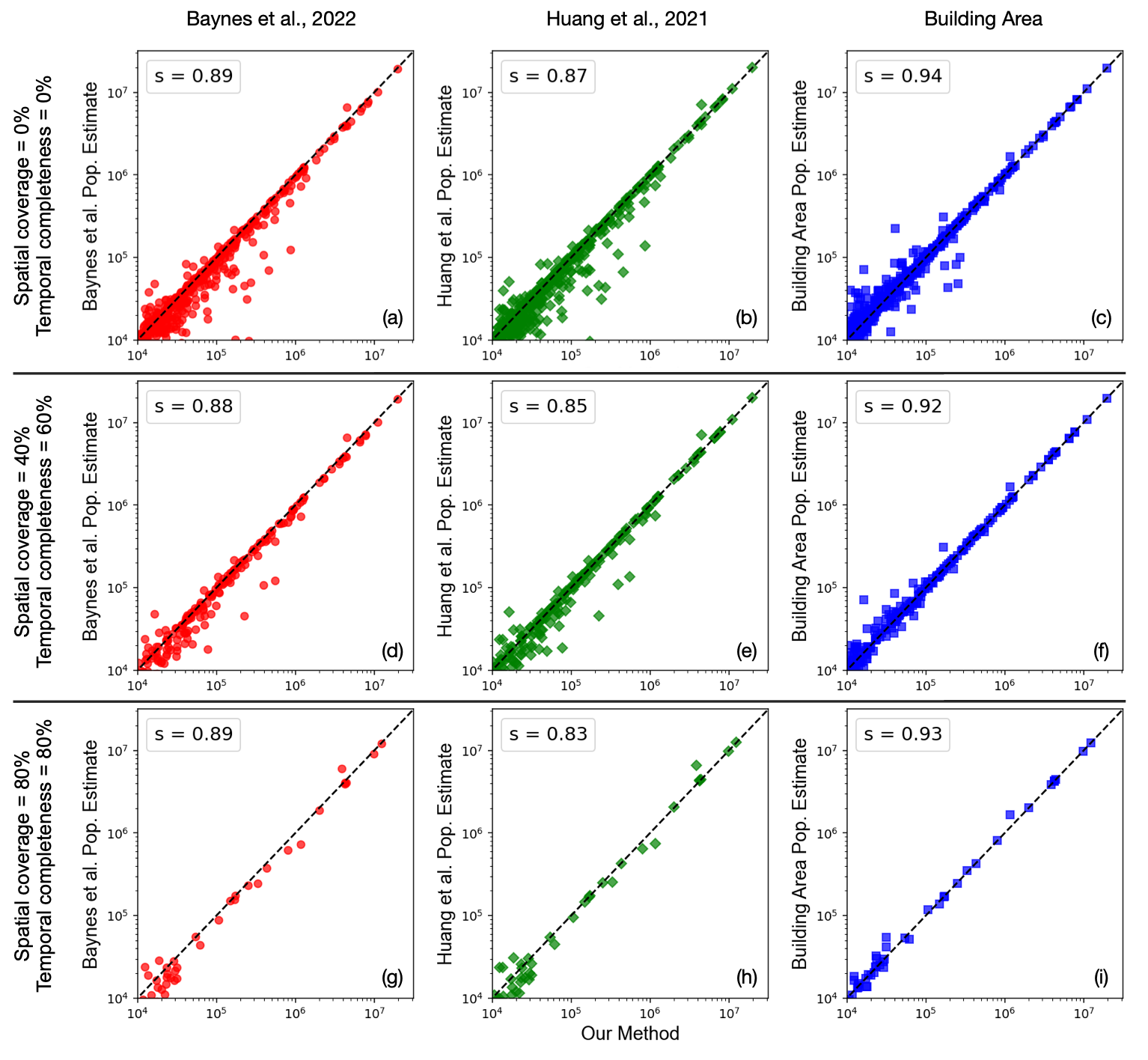}
    \caption{Comparisons between our population estimation method and competing methods for 2010-2015 populations. (a) Baynes et al. \cite{Baynes2022} city population estimates, (b) Huang et al. \cite{Huang2021} city population estimates, (c) building footprint-based city population estimates for temporal completeness and geospatial completeness $>0\%$. (d) Baynes et al. \cite{Baynes2022} city population estimates, (e) Huang et al. \cite{Huang2021} city population estimates, (f) building footprint-based city population estimates for temporal completeness $>60\%$ and geospatial completeness $>40\%$. (g) Baynes et al. \cite{Baynes2022} city population estimates, (h) Huang et al. \cite{Huang2021} city population estimates, (j) building footprint-based city population estimates for temporal completeness and geospatial completeness $>80\%$.  All city boundaries are as of 2015.
    }
    \label{fig:PopComparison_all}
\end{figure*}

\begin{figure*}[tbh!]
    \centering
    \includegraphics[width=0.8\linewidth]{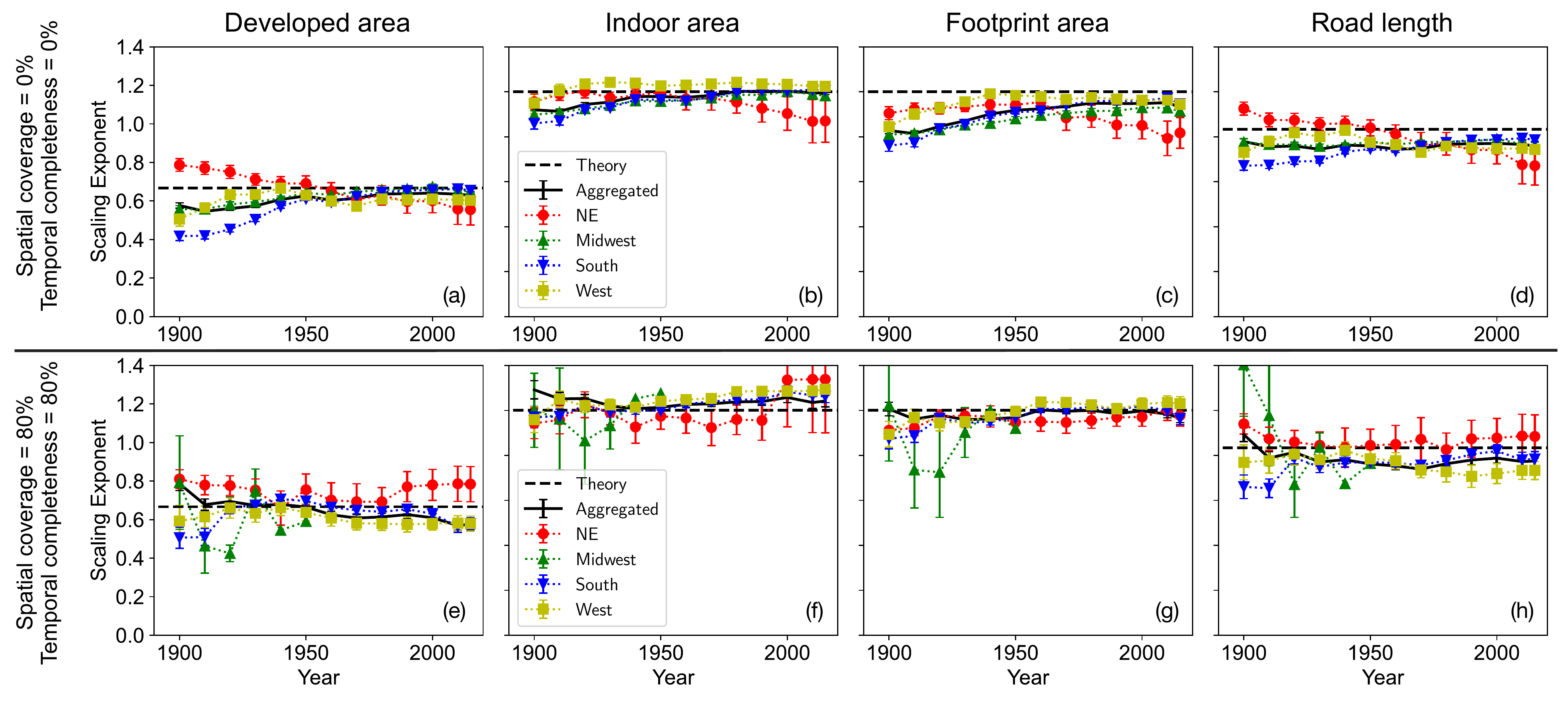}
    \caption{Scaling exponents over time. Plotted are exponents over time among all cities with CBSA temporal completeness and spatial coverage greater than 0\% and 80\%. Columns represent the city statistic: (a,e) developed area, (b,f) indoor area, (c,g) footprint area, and (d,h) road length. Data is separately analyzed by CONUS region: Northeast, Midwest, South, and West. Error bars represent standard errors.} 
    \label{fig:exp_00}
\end{figure*}

\begin{figure*}[tbh!]
    \centering
    \includegraphics[width=0.8\linewidth]{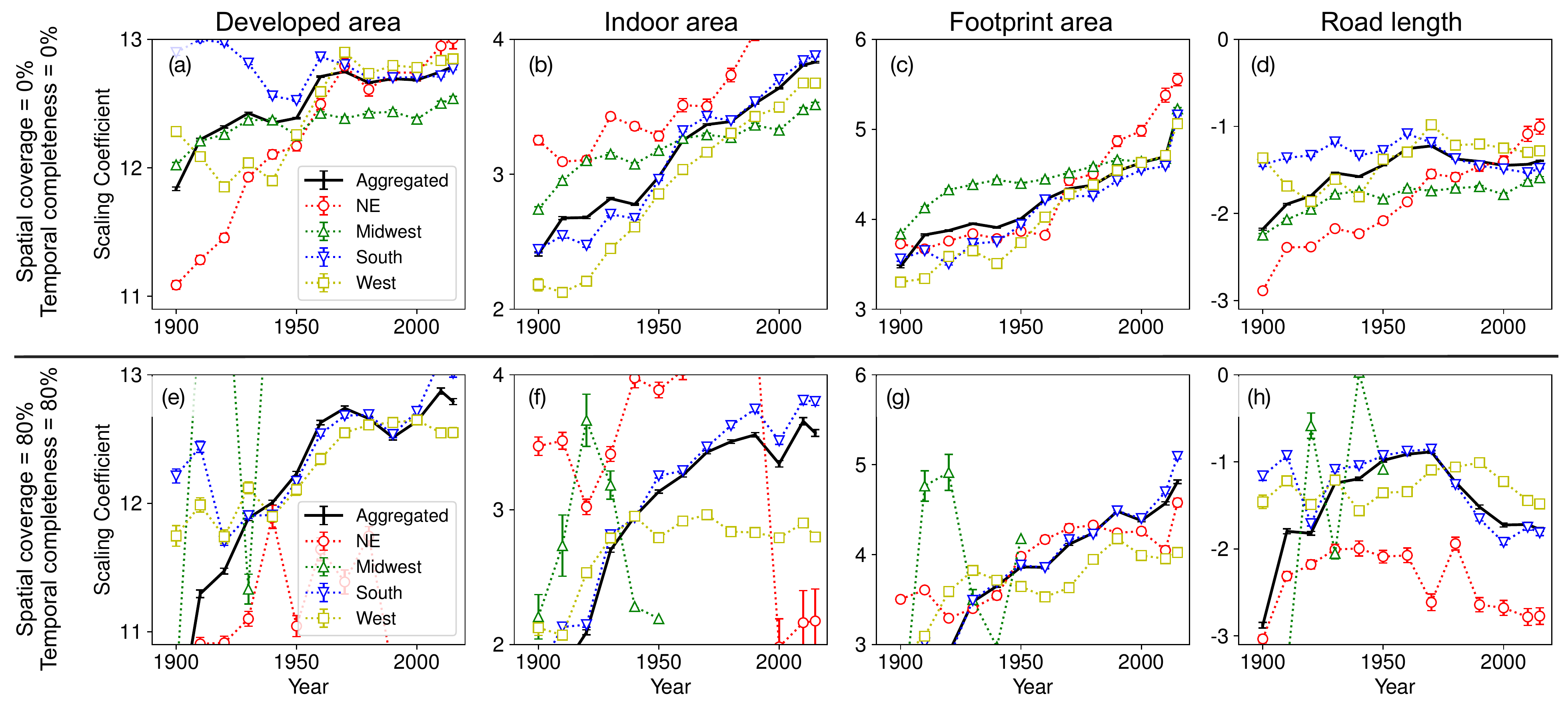}
    \caption{Scaling coefficients over time. Plotted are coefficients over time among all cities with CBSA temporal completeness and spatial coverage greater than 0\% and 80\%. Columns represent the city statistic: (a,e) developed area, (b,f) indoor area, (c,g) footprint area, and (d,h) road length. Data is separately analyzed by CONUS region: Northeast, Midwest, South, and West. Error bars represent standard errors.}
    \label{fig:coef_8080}
\end{figure*}

\begin{figure*}[tbh!]
    \centering
    \includegraphics[width=0.8\linewidth]{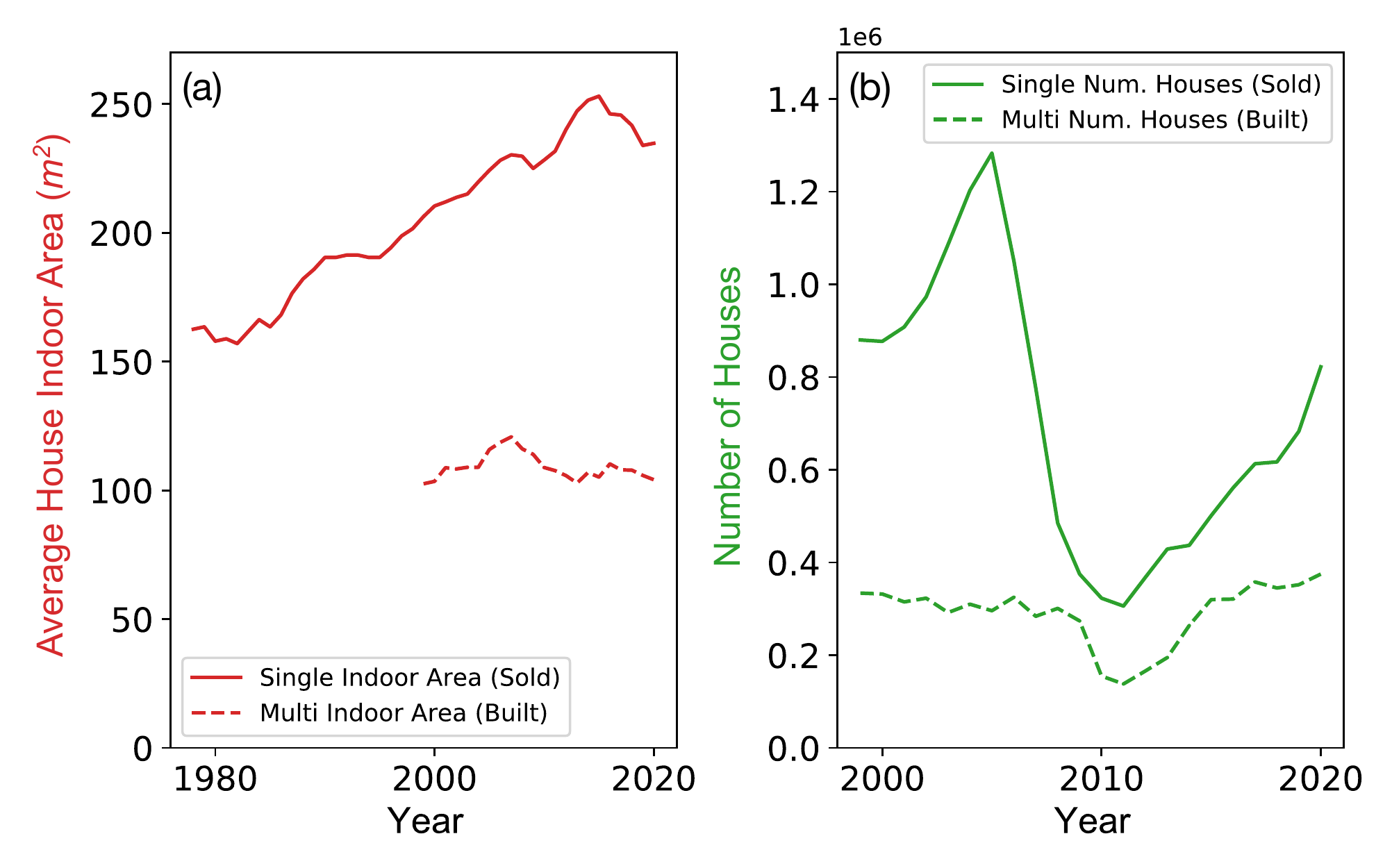}
    \caption{(a) Average house indoor area and (b) number of houses sold or built between 1978 and 2020. We discover that single family houses show substantial increases in their average indoor area over the past 40 years, while multi-family residences built, which are a smaller fraction of all residences, have a stable average area. The dip in houses sold around 2008 corresponds to a recession in the US. Data from \cite{HouseArea}. Number of multi-family houses sold are not reported in these data, while number of single-family houses built is not representative given the large number of houses left unsold after the 2008 US recession, but show similar trends.}
    \label{fig:house_area}
\end{figure*}

\begin{figure*}
    \centering
    \includegraphics[width=1\linewidth]{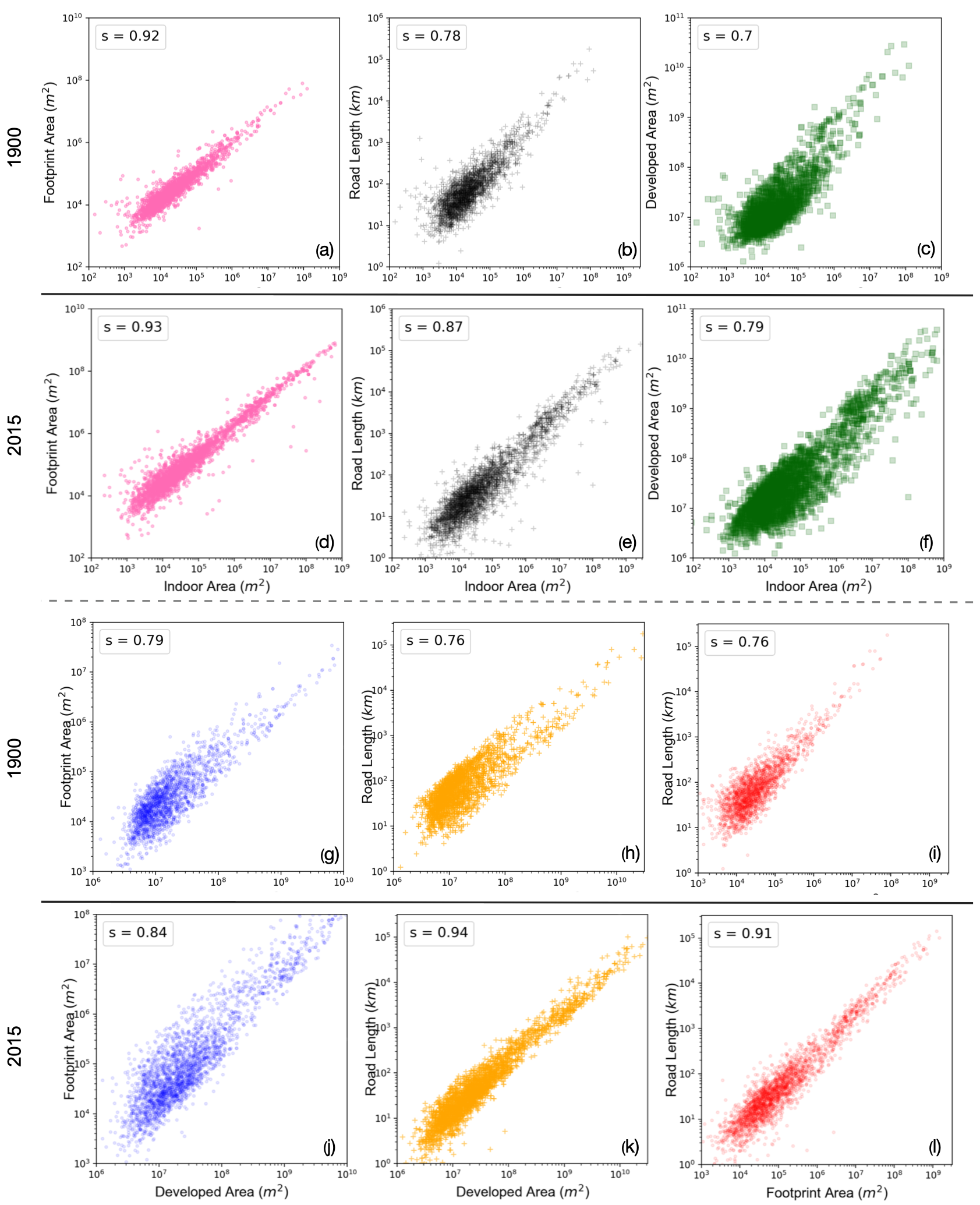}
    \caption{Relationship between different city features. (a,d) Footprint area versus indoor area, (b,e) road length versus indoor area, (c,f) developed area versus indoor area, (g,j) footprint area versus developed area, (h,k) road length versus developed area, (i,l) road length versus footprint area. Legends are the Spearman rank correlation. The first and third row shows features as of 1900 while the second and fourth row shows features as of 2015.}
    \label{fig:correlations_patch}
\end{figure*}

\clearpage

\end{document}